\documentclass{article}
\usepackage{emulateapj,apjfonts}

\def\deg{\ifmmode ^{\rm o} \else $^{\rm o}$\fi}

\def\arcsec{\ifmmode '' \else $''$\fi}
\def\arcmin{\ifmmode ' \else $'$\fi}
\def\arcsecpoint{\ifmmode ''\!. \else $''\!.$\fi}
\def\arcminpoint{\ifmmode '\!. \else $'\!.$\fi}
\def\kms{\ifmmode {\rm km\,s}^{-1} \else km\,s$^{-1}$\fi}
\def\Msun{\ifmmode M_{\odot} \else $M_{\odot}$\fi}
\def\Lsun{\ifmmode L_{\odot} \else $L_{\odot}$\fi}
\def\qo{\ifmmode q_{\rm o} \else $q_{\rm o}$\fi}
\def\Ho{\ifmmode H_{\rm o} \else $H_{\rm o}$\fi}
\def\ho{\ifmmode h_{\rm o} \else $h_{\rm o}$\fi}

\def\gtsim{\raisebox{-.5ex}{$\;\stackrel{>}{\sim}\;$}}
\def\vFWHM{\ifmmode v_{\mbox{\tiny FWHM}} \else
            $v_{\mbox{\tiny FWHM}}$\fi}
\def\CCF{\ifmmode F_{\it CCF} \else $F_{\it CCF}$\fi}
\def\ACF{\ifmmode F_{\it ACF} \else $F_{\it ACF}$\fi}
\def\Halpha{\ifmmode {\rm H}\alpha \else H$\alpha$\fi}
\def\Hbeta{\ifmmode {\rm H}\beta \else H$\beta$\fi}
\def\Hgamma{\ifmmode {\rm H}\gamma \else H$\gamma$\fi}
\def\Hdelta{\ifmmode {\rm H}\delta \else H$\delta$\fi}
\def\Lya{\ifmmode {\rm Ly}\alpha \else Ly$\alpha$\fi}
\def\Lyb{\ifmmode {\rm Ly}\beta \else Ly$\beta$\fi}

\def\ciii{\ifmmode {\rm C}\,{\sc iii} \else C\,{\sc iii}\fi}
\def\civ{\ifmmode {\rm C}\,{\sc iv} \else C\,{\sc iv}\fi}

\def\o5007{[O\,{\sc iii}]\,$\lambda5007$}

\def\caii{Ca\,{\sc ii}}

\def\fun#1#2{\lower3.6pt\vbox{\baselineskip0pt\lineskip.9pt
  \ialign{$\mathsurround=0pt#1\hfil##\hfil$\crcr#2\crcr\sim\crcr}}}
\def\lap{\mathrel{\mathpalette\fun <}}
\def\gap{\mathrel{\mathpalette\fun >}}

\def\MBH{\ifmmode M_{\bullet} \else $M_{\bullet}$\fi}
\def\Msigma{\ifmmode M_{\bullet}\,\mbox{--}\,\sigma 
	\else $M_{\bullet}$ -- $\sigma$\fi}

\slugcomment{Submitted to the Astrophysical Journal Letters}

\lefthead{Ferrarese et al.}
\righthead{Consistency of Black Hole Masses in Quiescent and Active Galaxies.}

\begin{document}

\title{Supermassive Black Holes in Active Galactic Nuclei. I. The
Consistency of Black Hole Masses in Quiescent and Active Galaxies.}

\author{Laura~Ferrarese,\altaffilmark{1,2}
Richard~W.~Pogge,\altaffilmark{2,3}
Bradley~M.~Peterson,\altaffilmark{2,3}
David~Merritt,\altaffilmark{1}
Amri~Wandel,\altaffilmark{4} and
Charles~L.~Joseph\,\altaffilmark{1}
}
\altaffiltext{1}{Department of Physics and Astronomy,
Rutgers University, New Brunswick, NJ 08854\\ {\tt lff@physics.rutgers.edu, 
merritt@physics.rutgers.edu, cjoseph@physics.rutgers.edu}}
\altaffiltext{2}{Visiting Astronomer, Kitt Peak National Observatory,
National Optical Astronomy Observatories, which is operated by the 
Association of Universities for Research in Astronomy, Inc. (AURA), 
under a cooperative agreement with the National Science Foundation.}
\altaffiltext{3}{Department of Astronomy, The Ohio State
University, 140  West 18th Avenue, Columbus, OH 43210\\ {\tt
peterson@astronomy.ohio-state.edu, pogge@astronomy.ohio-state.edu}}
\altaffiltext{4}{Racah Institute, Hebrew University,
Jerusalem 91904,  Israel\\ {\tt amri@frodo.fiz.huji.ac.il}}

\begin{abstract}
We report the first results of a program to measure accurate
stellar velocity dispersions in the bulges of the host galaxies of
active galactic nuclei (AGNs) for which  accurate black hole
(BH)  masses have been determined via reverberation mapping.  We find
good agreement between BH masses obtained from reverberation mapping,
and from the $M_{\bullet}-\sigma$ relation as defined by quiescent
galaxies, indicating a common relationship between active and
quiescent black holes and their larger-scale environments. 

\end{abstract}

\keywords{galaxies: active --- galaxies: Seyfert} 

\section{Introduction}

Two techniques have been successfully applied over the last few years
to derive masses of supermassive black holes (BHs) in samples of
galaxies.  Kinematics of stars or gas on scales of $\lap 1-10$ pc
yield estimates of \MBH\ based on Newton's laws under the assumption
that the observed motions reflect the gravitational influence of the
BH. This technique has been applied to galaxies near enough ($d\lap
100$ Mpc) that the radius of influence of the BH,
$r_h=GM_{\bullet}/\sigma_*^2$, is likely to have been resolved on the
angular scales accessible to ground- and space-based telescopes,
$1\arcsec \gap \theta \gap 0\arcsecpoint1$.  The second technique,
``reverberation mapping'' (RM), makes use of broad emission-line
variability to estimate BH masses in active galactic nuclei (AGNs).
The size of the broad emission-line region is estimated from time
delays between variations in the continuum (assumed to originate near
the BH) and the response of the broad emission lines (Blandford \&
McKee 1982; Peterson 1993; Netzer \& Peterson 1997).  By combining the
time delays with the emission-line widths, the BH mass then follows
from the virial theorem.  Black hole masses based on RM have been
published for about three dozen AGNs (Wandel, Peterson \& Malkan 1999;
Kaspi et al.\ 2000).

Both mass-estimation techniques have strengths and weaknesses.  Masses
derived from RM are based on gas that is much closer to the BH, often
$r\lap 0.01$ pc, than the stars or gas used in kinematical studies. 
At these separations, the gravitational force to which the gas
responds is due almost entirely to the BH and there is little danger
of a ``false detection.''  However the precise relation between \MBH\
and the measured quantities is uncertain at levels of $\sim 50$\% due
primarily to uncertainties in the geometry and kinematics of the BLR and
other unknowns (e.g., Krolik 2001).

In the case of BH masses derived from spatially resolved kinematics,
formal accuracies can be high provided $r_h$ has actually been resolved;
otherwise, estimates of \MBH\ depend sensitively on assumptions
about the character of the motion near the BH. Here the danger is that
the inferred BH mass is actually a measure of the {\it combined} mass
of the BH and the stars in the resolved region, $r\gg r_h$; modeling
techniques applied to such data can substantially overestimate \MBH\
by ascribing too much of the gravitational force in the nucleus to the
BH and too little to the stars.

In fact, there have been claims of systematic differences between RM
and kinematical BH masses, in the sense that RM masses at a given
bulge luminosity, $M_B$, are a factor $\sim 20$ smaller than
kinematical  masses (Wandel 1999). The discrepancy has most often been
ascribed to systematic errors  in the RM masses (Richstone et al.
1998; Faber 1999; Ho 1999).  Unfortunately, it is extremely difficult
to make a direct comparison between dynamical and reverberation-based
BH masses.  In AGNs, the variable nonstellar continuum and broad
emission lines that enable RM tend to swamp the stellar features at $r
< r_h$ which are needed to study the stellar dynamics.  At the present
time, there is no galaxy for which the BH mass has been measured using
both techniques.

A possible solution to this problem emerged with the discovery of the
\Msigma\ relation, a tight empirical correlation between \MBH\ and the
stellar velocity dispersion $\sigma$ (Ferrarese \& Merritt 2000,
hereafter FM00; Gebhardt et al.  2000a, hereafter G00a).  The latter
is defined on scales much larger than $r_h$ and hence is easily
accessible to ground-based instruments.  Since stellar velocity
dispersions can also be measured in active galaxy hosts, it is
possible to determine whether the \Msigma\ relation for AGNs is
similar to that of quiescent galaxies.  Preliminary attempts at such a
comparison have been published (Gebhardt et al.\ 2000b [hereafter
G00b]; Merritt \& Ferrarese 2001a [hereafter MF01a]) based on
published AGN bulge velocity dispersions (principally from Nelson \&
Whittle 1995, hereafter NW95).  These studies reveal a general
consistency between the AGN data and the \Msigma\ relation for normal
galaxies, confirming that the apparent discrepancies between the
kinematic and RM masses were due mostly to systematic overestimates in
the BH masses derived from stellar kinematic measurements (MF01a,
Merritt \& Ferrarese 2001b).  Unfortunately, the uncertainties in both
the AGN BH masses and stellar velocity dispersions used by G00b and
MF01a are too large to provide a truly critical test.  The reason for
the large uncertainties in \MBH\ is quite simple: the largest source
of random errors in AGN BH masses is the light-travel time delay for
the emission lines, which determines the size of the line-emitting
region.  The fractional error in this quantity is largest for those
sources that have the smallest line-emitting regions, since the
uncertainties are determined in large part by the time resolution of
the monitoring program.  The smallest line-emitting regions are found
in the least-luminous objects (i.e., $R \propto L^{0.5-0.7}$; Kaspi et
al.\ 2000).  Generally speaking, lower-luminosity AGNs have not been
monitored any more closely than higher-luminosity AGNs and thus the
fractional errors in their BH masses are relatively large, in some
cases as large as 100\%.  The low-luminosity AGNs are the very
galaxies for which bulge velocity dispersions have been measured, and
thus do not yet provide a conclusive argument for consistency between
the \Msigma\ relations in AGNs and normal galaxies.

In order to better compare the \Msigma\ relations for AGNs and 
quiescent galaxies, we are undertaking a program to measure bulge
velocity dispersions of the host galaxies of AGNs with formally
well-determined reverberation-based BH masses.  In this contribution,
we describe the first results from this program.

\begin{deluxetable}{lcccccrr}
\tablecolumns{7}  
\tablecaption{AGN Sample Observed}  
\tablewidth{0pt}
\tablehead{  
\colhead{Galaxy} &  
\colhead{Redshift} &  
\colhead{\MBH} &
\colhead{Source\tablenotemark{1}} &   
\colhead{Position} &
\colhead{Exposure Time} &  
\colhead{$\sigma$(FCQ)} &
\colhead{$\sigma$(MPL)} \nl  
\colhead{ } & \colhead{$z$} &
\colhead{($10^7 \Msun$)} &   
\colhead{ } & \colhead{Angle} &
\colhead{(sec)}&  
\colhead{(\kms)} & 
\colhead{(\kms)} \nl
\colhead{(1)} &  
\colhead{(2)} &  
\colhead{(3)} &   
\colhead{(4)} &
\colhead{(5)} & 
\colhead{(6)} &  
\colhead{(7)} &
\colhead{(8)} }  
\startdata  
NGC 4051&0.0024&$0.14^{+0.10}_{-0.06}$ &1&$135\deg$& 3600 & $80\pm4$ & $80\pm3$\nl  
NGC 4151&0.0033&$1.20^{+0.83}_{-0.70}$    &2&$135\deg$&3600 & $93\pm5$ & $85\pm5$\nl  
NGC 5548&0.017 &$5.9\pm2.5$ &3&$110\deg$&7200 &  $183\pm10$ &$180\pm6$ \nl  
Mrk 79  &0.022&$10.2^{+3.9}_{-5.6} $  &2&$56\deg$ &7200 & $130\pm9$ & $120\pm8$ \nl
Mrk 110 &0.036 &$0.77^{+0.38}_{-0.29}$  &2&$110\deg$&15600& $86\pm5$ & $95\pm8$ \nl  
Mrk 817 &0.031 &$3.54^{+1.03}_{-0.86}$ &2&$90\deg$ &12000& $142\pm6$ & $140\pm8$ \nl  \enddata
\tablenotetext{1}{Data source for masses: 1: Peterson et al.\
(2000); 2: Kaspi et al.\ (2000); 3: Peterson \& Wandel (2000)}
\end{deluxetable}

\section{Observations and Data Reduction}

\subsection{Sample Selection}

Black hole masses based on reverberation studies have been published
for 34 AGNs (Wandel, Peterson \& Malkan 1999; Kaspi et al.\ 2000).
There are both systematic and random uncertainties in these
measurements: systematic errors arise primarily from the largely
unknown geometry and kinematics of the line-emitting region, and the
random errors are due mostly to uncertainties in the size of the
line-emitting region, as noted earlier, and the uncertainty in the
translation of the observed FWHM  to the 3-D velocity dispersion.  The
typical level of random uncertainty in the reverberation-based BH
masses is $\gtsim50$\%.  We focus our attention on those AGNs with
better-determined BH masses (i.e., smaller random errors). These
provide the most meaningful comparison with normal galaxies: the
scatter in the \Msigma\ relation for normal galaxies is $\sim30$\%,
while the intrinsic scatter in the relation appears to be negligible
(FM00).

\subsection{Observations}
We have obtained bulge velocity dispersions from the CaII triplet
lines at rest wavelengths 8498, 8542, and 8662\,\AA. These lines are
in the spectral region where the AGN contribution is minimized (e.g.,
NW95).

The objects observed in this study are listed in Table 1.  Column (1)
gives the common name of the object, and its redshift is given in
column (2).  Column (3) gives the reverberation-based virial mass of
the central BH in each object, and the published source of the virial
mass is given in column (4).

The observations reported here were obtained on UTC 2001 April 7--9 on
the Mayall 4-m Telescope of the National Optical Astronomy
Observatories on Kitt Peak.  The observations were made with the
Ritchey-Chretien Spectrograph with the BL380 grating (1200 lines per
mm, blazed at 9000\,\AA, centered at approximately 8910\,\AA), and an
RG610 blocking filter.  The detector employed was the T2KB CCD, a
back-illuminated $2048 \times 2048$ device windowed to $440 \times
2048$ pixels.  The slit width was set to 2\arcsec.  The resulting spectra
have a resolution $R \approx 5000$, and cover the range
8150--9660\,\AA. To allow accurate sky-subtraction, a long slit
($\sim5\arcmin$) was used.  Column (5) of Table 1 gives the
spectrograph position angle at which the observations were made; when
such information was available, we aligned the slit along the major
axis of the galaxy.  Column (6) gives the total integration time on
each source.

Each observation of a galaxy was bracketed with quartz-lamp
spectra for flat-field correction, and He-Ne-Ar spectra for wavelength
calibration.  A spectrum of the twilight sky was obtained to determine
a nightly cross-dispersion illumination correction.  We also obtained
spectra of 14 late-type giant stars (G8 III --- K6 III) that were used
as templates and radial-velocity standards in the analysis described
below.

\subsection{Data Reduction}

The raw long-slit spectra were reduced by flat-field, bias, and
slit-illumination corrections following the procedures described by
Pogge (1992).  We used the {\sc XVista}\footnote{See
http://ganymede.nmsu.edu/holtz/xvista.} package for all reductions.
Curvature of the long-slit spectra in the dispersion and
cross-dispersion directions was removed by tracing stars and bright,
isolated night-sky emission lines in the spectra and rectifying the
long-slit spectra.  In this spectral region, the T2KB CCD is subject
to fringing at the 10\% level longward of $\sim$9000\,\AA, which we
reduced by using the bracketing flat-field exposures.  In general,
residual fringing was less than 0.1\% in the spectral regions of
interest.  We combined the longslit spectra in groups of three 20-min
exposures to remove cosmic-ray events by sigma clipping with a simple
CCD noise model (gain = 1.9 e$^-$, $\sigma_{\rm read-out}=5$e$^-$).
The nuclear spectra were extracted from each group within a $2\arcsec
\times 4\arcsec$ aperture centered on the nucleus.  For the fainter
objects, we used a 4\arcsecpoint8 extraction aperture to improve the
signal-to-noise ratio.  Sky spectra were extracted from flanking
regions along the slit and subtracted from the nuclear spectra.  We
then combined all of the nuclear spectra into a final spectrum.  This
group-wise extraction of spectra taken within at most an hour of each
other reduced the effects of variability in the atmospheric OH airglow
lines. The fully reduced spectra of all objects are shown in Fig. 1.

We point out that the \Msigma\ relation was defined by FM00 using
central velocity dispersions corrected to an aperture of size $r_e/8$,
with $r_e$ the effective radius of the galaxy.  G00a, on the other
hand, used spatially averaged, rms, line-of-sight velocity dispersion
within $r_e$.  MF01a demonstrated that there is no systematic
difference between the two definitions of $\sigma$ in this sample.
The bulge effective radius is not known for any of our target
galaxies, but the bulge is enclosed within the aperture used for the
spectral extraction.  On this basis, in \S 3 we treat
our measured dispersions on the same basis as the aperture-corrected
dispersions tabulated by MF01a.

\subsection{Data Analysis}

Two independent estimates of the stellar velocity dispersion were
obtained from the \caii\ triplet.  The first employed the Fourier
Correlation Quotient (FCQ) method described by Bender (1990) and
Bender, Saglia, and Gerhard (1994).  The second was based on the
Maximum Penalized Likelihood (MPL) method developed by Merritt (1997). 
FCQ and MPL differ in 

\centerline{\includegraphics[width=6.4cm]{fig1.epsi}}
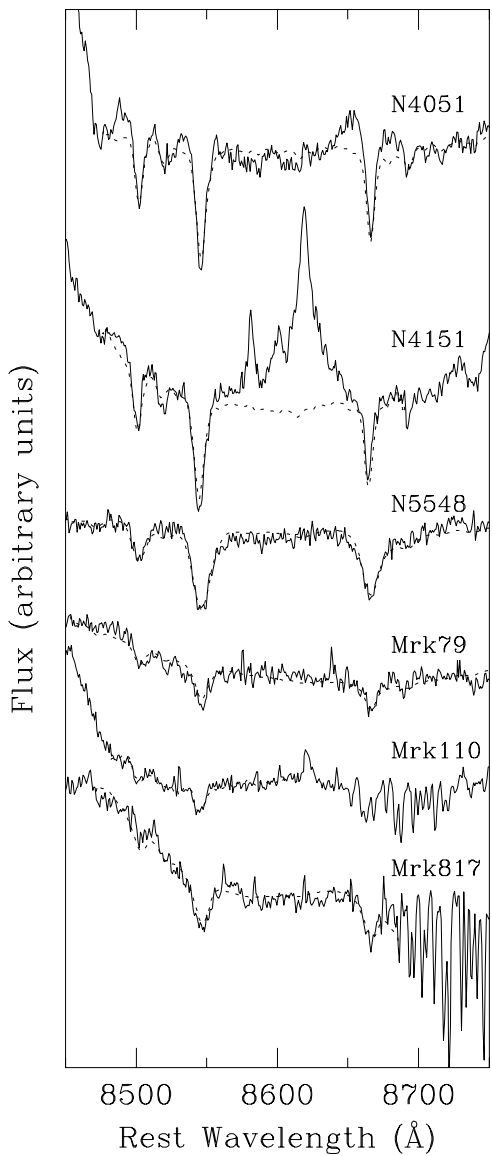
\figcaption{Calibrated spectra for the six galaxies in our
program.  An arbitrary additive constant has been added to the fluxes
on the vertical axis.  The dotted line represents the best fit
returned by the FCQ method.}
\centerline{}

\vskip -.2in
\noindent important ways (e.g., Joseph et al.\ 2001);
using both allows us to assess the impact of systematics on the
measured values.  MPL was applied to the fully calibrated and
continuum normalized spectra.  FCQ incorporates a continuum fitting
routine, thus was applied to the spectra before normalization. 
The velocity dispersion measurements resulting from FCQ and MPL are
listed in columns (7) and (8) of Table 1, respectively.  The errors
associated with $\sigma$ are the formal random uncertainties returned
by the FCQ and MPL codes, but there are likely unaccounted systematic
uncertainties.  Repeating the analysis using different stellar
templates, or changing the parameters for the continuum fits in FCQ
show that a more realistic error estimate is of order 15\%.

\section{Results and Discussion}

No velocity dispersions have been previously published for Mrk79,
Mrk110 and Mrk817.  NW95 listed velocity dispersions from CaII and/or
Mgb absorption lines for six galaxies with reverberation-based BH
masses, including NGC4151 and NGC 5548.  Unfortunately, NW95 warn that
only for two of these galaxies, NGC 4051 and 3C120, are the measured
dispersions deemed accurate.  For NGC 4051, they quote $\sigma = 88
\pm 13$ \kms, in good agreement with our result.  For 3C120, which has
a RM mass of $3.0^{+1.9}_{-1.4}\times10^7\,\Msun$ (Wandel, Peterson \&
Malkan 1999; Kaspi et al.\ 2000) NW95 list $\sigma = 162 \pm 20$ \kms
(from Smith, Heckman, \& Illingworth 1990). For lack of an
alternative, the NW95 velocity dispersions for all six galaxies  were
adopted by G00b and MF01a, in spite of their large associated
uncertainties.

In Fig.\ 2, we show the relationship between BH
mass and bulge velocity dispersion (FCQ values with 15\% errors
assumed) for the six galaxies listed in Table 1, along with the same
relationship for quiescent or weakly active galaxies (full references
for both \MBH\ and $\sigma$ for the latter can be found in MF01a).
For completeness, we also plot, as small open circles, the sample of
galaxies for which values of \MBH\ were tabulated by G00a based on
dynamical modeling of {\em Hubble Space Telescope} data, although we
refrain from using these data in the following discussion since the
modeling and analysis of the data that led to those \MBH\ measurements
have yet to be published.

While secure conclusions are premature given the small sample size,
several points are worth mentioning.  As noted by G00b and MF01a based
on lower-quality data than presented in this paper, the current
evidence is that reverberation-based masses are not systematically
underestimated as suggested by, e.g., Richstone et al.\ (1998), Faber
(1999) and Ho (1999).  We stress that the data presented in this paper
are unique for having high-quality measurements of both \MBH\ and
$\sigma$.  Our data support the conclusion that dynamical and
reverberation-based masses are generally consistent, and that the
two methods can potentially yield BH mass estimates of comparable
precision.  This is of particular relevance.  Reverberation mapping is
intrinsically unbiased with respect to BH mass, provided the galaxies
can be monitored at closely spaced time intervals.  While dynamical
methods rely on the ability to spatially resolve the region dominated
by the BH gravitational potential, RM samples a region which is per se
unresolvable.  It follows that an aggressive monitoring campaign of
AGNs is the only viable method to probe the low-mass (\MBH\ less than
a few million solar masses) end of the \Msigma\ relation.  It is also
apparently as good a method as traditional dynamical studies to probe the
higher-mass regime, although the obvious drawback is that it is
perforce limited to galaxies with Type 1 AGN, which constitute only
$\sim1$\% of the general galaxy population.  Probing the low mass end
of the \Msigma\ relation is of particular interest since the slope and
scatter of the relation have important implications for both
hierarchical models of galaxy formation (Haehnelt, Natarajan, \& Rees
1998; Silk \& Rees 1998; Haehnelt \& Kauffmann 2000) and the effect of
mergers on the subsequent evolution (Cattaneo et al.\ 1999), including
the coalescence of the binary BHs which are expected to form as a
result of galaxy mergers (Milosavljevic \& Merritt 2001).

Furthermore, RM can probe galaxies at high redshift and with a wide
range of nuclear activity, opening an avenue to explore possible
dependences of the \Msigma\ relation on redshift and activity level. 
Regarding the latter point, we note that NGC~4051, the galaxy with the
smallest \MBH\ in our sample, is a narrow-line Seyfert 1 (NLS1)
galaxy.  Mrk~110 also approximately meets the traditional narrow-line
Seyfert 1 criterion that ${\rm FWHM}(\Hbeta) \leq 2000$\,\kms.  The BH
masses of the NLS1s are consistent with those of the other AGNs, and
all are consistent with the masses inferred from the \Msigma\ relation
defined by the quiescent galaxies.  This result is at odds with the
recent findings of Mathur et al.  (2001), who claim that NLS1s lie
systematically below the \Msigma\ relation defined by Seyfert 1
galaxies.  One paradigm for NLS1s is that, compared to AGNs with
similar non-thermal luminosity, they are powered by lower mass BHs
accreting at a relatively higher accretion rate.  A competing
explanation is that they differ from other AGNs only in inclination;
the small widths of the broad lines and concomitantly low
reverberation masses are due to viewing disk-like sources at low
inclination (i.e., nearly face-on).  The consistency of the BH mass
and bulge velocity dispersion of NGC~4051, in particular, with the
\Msigma\ relation for other galaxies strongly favors the low-mass/high
accretion rate interpretation for the nature of NLS1s.

A linear fit to the sample of 18 galaxies --- 12 galaxies from
FM00 plus the six reverberation-mapped galaxies
--- using regression with bivariate errors and intrinsic scatter
(Akritas \& Bershady 1996; see also discussion in MF01a) gives a slope
of 

\centerline{\includegraphics[width=8.0cm]{fig2.epsi}}
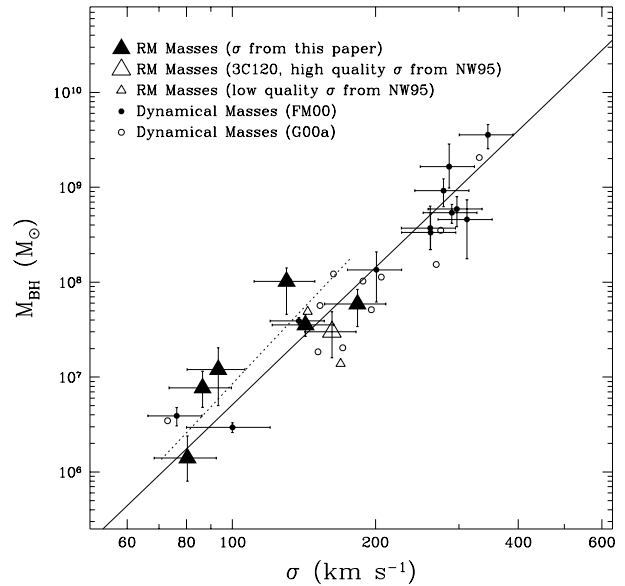
\figcaption{Black hole mass versus central velocity dispersion
$\sigma$ of the host elliptical galaxy or bulge.  Small filled circles
and errorbars represent quiescent or weakly active galaxies with
dynamical measurements of \MBH. Open circles represent unpublished
dynamical BH masses listed by G00a.  Large solid triangles are the
reverberation-mapped galaxies presented in this paper, while open
triangles are reverberation-mapped galaxies for which $\sigma$ is
available in the literature.  The solid line is the best linear fit to
the \Msigma\ relation as published by MF01a for the kinematical masses
only (filled circles), with slope $4.81\pm0.55$.  The dotted line is
the best fit to the six RM galaxies presented in this paper
(filled triangles), and has a slope $5.3\pm1.6$.}
\centerline{}
\vskip .3in

\noindent $4.50 \pm 0.34$, consistent with the slope of $4.81 \pm 0.55$
derived by MF01a using only the kinematical mass estimates. 
To
conclude, we note that three of the galaxies in Fig.\ 2, NGC~4151,
Mrk~79, and Mrk~110, seem to lie slightly {\it above} the ``fiducial''
\Msigma\ relation defined by MF01a; i.e., these galaxies have black
holes larger than expected from their bulge velocity dispersions. 
This is unlikely to be attributable to systematic errors in $\sigma$,
since the quality of the fit returned by both FCQ and MPL for these
galaxies, while not as good as for NGC~4051 and NGC~5548, is
comparable to that of Mrk~817, which lies on the fiducial relation. 
There also seems to be no correlation between residuals and
distance\footnote{Note that reverberation-mapping masses are
independent of distance, therefore the scatter cannot be due to errors
in the distance determination to these galaxies.}.  A more detailed
discussion is premature at this point: the small discrepancy could be
due to small systematic errors in the RM masses, or, perhaps more
interesting, may indicate that the current level of activity has
allowed for significant growth of the BH mass.  These results 
underscore the importance of further testing the \Msigma\ relation by measuring
$\sigma$ for a larger sample of reverberation-mapped galaxies over a
large BH mass range.

\acknowledgements{We are grateful to NOAO for observing time on the
4-m telescope, and especially to H.~Halbedel and B.~Gillespie for
capable assistance at the telescope.  LF acknowledges NASA LTSA grant
NAG5-8693. This research has made use of the NASA/IPAC Extragalactic
Database (NED), which is operated by the Jet Propulsion Laboratory,
California Institute of Technology, under contract with the National
Aeronautics and Space Administration.}

\end{document}